# Support Vector Machine For Transient Stability Assessment: A Review

Umair Shahzad
Department of Electrical and Computer Engineering,
University of Nebraska-Lincoln,
Lincoln, NE, USA
umair.shahzad@huskers.unl.edu

*Abstract*–Accurate transient stability assessment is a crucial prerequisite for proper power system operation and planning with various operational constraints. Transient stability assessment of modern power systems is becoming very challenging due to rising uncertainty and continuous integration of renewable energy generation. The stringent requirements of very high accuracy and fast computation speed has further necessitated accurate transient stability assessment for power system planning and operation. The traditional approaches are unable to fulfil these requirements due to their shortcomings. In this regard, the popularity of prospective approaches based on big data and machine learning, such as support vector machine, is constantly on the rise as they have all the features required to fulfil important criteria for real-time TSA. Therefore, this paper aims to review the application of support vector machine for transient stability assessment of power systems. It is believed that this work will provide a solid foundation for researchers in the domain of machine learning and computational intelligence-based applications to power system stability and operation.

TABLE I. TABLE OF ABBREVIATIONS

| Abbreviation | Meaning |
|---|---|
| ANN | Artificial Neural Network |
| ASVM | Aggressive SVM |
| BP | Back Propagation |
| CSVM | Conservative SVM |
| CVM | Core Vector Machine |
| DL | Deep Learning |
| DSA | Dynamic Security Assessment |
| DT | Decision Tree |
| EAC | Equal Area Criterion |
| EEAC | Extended Equal Area Criterion |
| FCT | Fault Clearing Time |
| GA | Genetic Algorithm |
| IoT | Internet of Things |
| LS-SVM | Least Squares SVM |
| ML | Machine Learning |
| MLP | Multi-layer Perceptron |
| ML-SVM | Multi-layer SVM |
| PMU | Phasor Measurement Unit |
| PNN | Probabilistic Neural Network |
| PTS | Probabilistic Transient Stability |
| RBF | Radial Basis Function |
| RFE | Recursive Feature Elimination |
| RTDS | Real Time Digital Simulator |
| SAE | Stacked Automatic Encoder |
| SLT | Statistical Learning Theory |
| SOFM | Self-Organizing Feature Map |
| SVM | Support Vector Machine |
| SVR | Support Vector Regression |
| TEF | Transient Energy Function |
| TSA | Transient Stability Assessment |
| WAMS | Wide Area Measurement System |

## I. INTRODUCTION

The growing load demand in power systems without accompanying investments in generation and transmission has impacted the evaluation of transient stability, demanding more reliable and faster tools. One of the most intriguing challenges in online operation of power systems is the evaluation of transient stability. Its significance has intensified due to the reduction of operational safety margins, increasing renewable generation and the introduction of competitive electricity market. Conventional analytical techniques such as time domain simulation and direct approaches do not allow to take preventive or corrective actions in appropriate time.

The most common and simplest approach to compute the transient stability status of a power system is the time-domain simulation of the nonlinear differential equations which govern the power system [1]. This approach requires accurate information on the system topology during and after the disturbance and thereby, it is time consuming. Another method for determining the stability after a contingency is the Transient Energy Function (TEF) approaches based on Lyapunov stability or Energy Function theory [2]. In this technique, the stability assessment is performed by comparing the difference between the kinetic energy and potential energy against a reference value for a specific fault. However, there are obstacles in computing the levels of kinetic and potential Energy under particular disturbances for large scale power systems [2]. The Equal Area Criterion (EAC) approach is based on the same theory and offers a way to evaluate the transient stability of a multimachine system represented as a one machine connected to an infinite bus system without solving the cumbersome system of differential-algebraic equations. Although EAC is powerful graphic approach, it involves obtaining an equivalent machine and allows only the classical generator model that represents only the generator's mechanical dynamics [3]. The Extended Equal Area Criterion (EEAC) is a fusion of the time-domain simulation and energy functions [4]. Although, this approach is computationally more efficient, but less accurate than time-domain simulation. A potential solution to conquer the flaws of the above-mentioned approaches for transient stability assessment (TSA) is the application of novel soft computing approaches [5].

In recent times, various different Machine Learning (ML) approaches have been proposed for real-time TSA. For instance, Decision Tree (DT) [6-7] is one of the pragmatic algorithms for predicting power system transient stability. Moreover, ensemble DT (Random Forest) has been considered for transient stability evaluation [8]. Artificial Neural Networks (ANNs) have also been considered to enhance the performance of transient stability status prediction [9-10]. Support Vector Machine (SVM) is regarded one of the most useful approaches used in real-time TSA [11]. As mentioned there are various ML approaches for TSA; however, this paper specifically focuses on SVM.

The rest of the paper is organized as follows. Section II discusses background and overview of ML. Section III elaborate various steps of ML. Section IV provides background and overview of SVM. Section V and VI briefly describes types and kernel function associated with SVM, respectively. Section VII provides a summary of various work of SVM application to TSA. Section VIII provides research gaps and suggestions for future work. Finally, Section IX concludes the paper.

## II. MACHINE LEARNING: OVERVIEW AND BACKGROUND

ML is broadly regarded as the subset of artificial intelligence [12] (simulation of human intelligence in machines, which are programmed to think like humans and mimic their actions), as outlined by Fig. 1. ML basically is an application of artificial intelligence that provides systems the ability to automatically learn and enhance from experience without being explicitly programmed [12-13]. In fact, the ML performs data analysis, using a set of instructions, through a variety of algorithms, for decision making and/or predictions [14]. Laborious designing and programming of algorithms are essential to be conducted, for ML, to implement diverse functionalities, such as, classification, clustering, and regression. Deep Learning (DL) is a class of ML algorithms that uses multiple layers to progressively extract higher-level features from the raw input. For instance, in image processing, lower layers may identify edges, whereas higher layers may distinguish the concepts relevant to a human being, such as digits, letters or faces [15]. It is majorly used for speech recognition, computer vision (high-level understanding from digital images or videos), medical image analysis, and natural language processing. There are numerous architectures used in DL such as deep neural networks, deep belief networks, recurrent neural networks, long short-term memory, and convolutional neural networks. The DL generally requires huge processing power and massive data [15]. The focus of this work is, however, on ML.

ML differs from traditional programming, in a very distinct manner. In traditional programming, the input data and a well written and tested program is fed into a machine to produce output. When it comes to ML, input data along with the output is fed into the machine during the learning phase, and it works out a program for itself. This is illustrated in Fig. 2 [16]. During the last decade, ML, and DL has demonstrated promising contributions to many research and engineering areas, such as data mining [17], medical imaging [18], communication [19], multimedia [20], geoscience [21], remote sensing classification [22], real-time object tracking [23], computer vision-based fault detection [24], and so forth. The integration of advanced information and communication technologies, specifically Internet of Things (IoT), in the power grid infrastructures, is one of the main steps towards the smart grid. Since the vital capability of IoT devices is their capability to communicate data to other devices in a more pervasive fashion, and hence a massive amount of data is made available at the control centers. Such meaningfully enhanced system condition awareness and data availability necessitates ML-based solutions and tools to conduct efficient data processing and analysis, to encourage the system operational management and decision-making [25]. Therefore, ML has been applied in various fields of power system, such as load forecasting [26], fault diagnosis [27], substation monitoring [28], reactive power control [29], unit commitment [30], maintenance scheduling [31], wind power prediction [32], energy management [33], load restoration [34], solar power prediction [35], state estimation [36], TSA [37], economic dispatch [38], and electricity price forecasting [39].

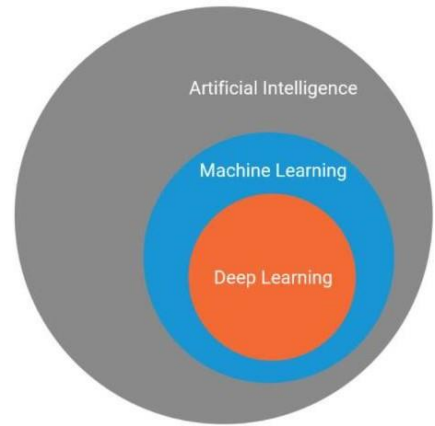

Figure 1. ML as a subfield of artificial intelligence

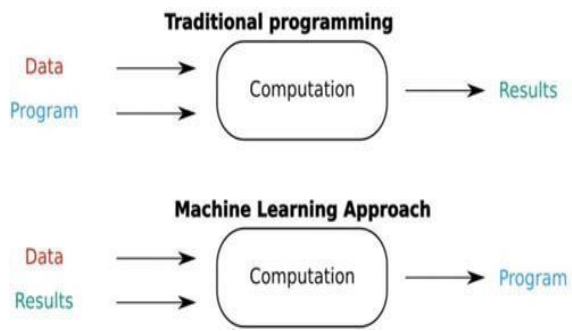

Figure 2. Traditional programming vs. ML

## III. STEPS OF MACHINE LEARNING

There are seven main steps of successfully implementing ML. They are outlined below and illustrated in Fig. 3 [40].

*1) Gathering Data*

The first and the most significant step of ML is gathering data. This step is very critical, as the quality and quantity of data gathered will directly determine how good the predictive model will turn out to be. The data collected is then tabulated and is commonly called as the training or learning data.

*2) Data Preparation*

After the training data is gathered, the next step of ML is data preparation, where the data is loaded into a suitable place and then, prepared for use in ML training. Here, the data is first put all together and consequently, the order is randomized as the order of data should not affect what is learned. This is also a good chance to do any visualizations of the data, as this will help see if there are any pertinent relationships between the different variables, and presence of any data imbalances or anomalies. Also, at this stage, the data must be divided into two parts. The first part, which is used in training the model, will be most of the dataset and the second will be used for the evaluation (validation and testing) of the performance of the trained model.

*3) Model Selection*

The subsequent step that follows in the workflow is choosing a model among the many that researchers and data scientists have created over the years. There are different algorithms for different tasks. Some are appropriate for image data, others for sequences (such as text, or music), some for numerical data, others for text-based data. A selection should be made based on the task required.

*4) Training*

After the above-mentioned steps are completed, the next step involves training, where the data is used to incrementally improve the ability of the model to predict. The training process requires initializing some random values for the model, predicting the output with those values, then comparing it with the model's prediction and eventually, adjusting the values such that they match the predictions that were made formerly. This process then replicates, and each cycle of updating is called one training step.

*5) Evaluation*

Once training is complete, evaluation is performed. This is where the testing dataset comes into play. Evaluation allows the testing of the model against data that has never been seen and used for training and is meant to be illustrative of how the model might perform in the real world.

*6) Hyperparameter Tuning*

Once the evaluation is over, any further improvement in the training process is possible by tuning the parameters. There were a few parameters that were implicitly assumed when the training was done. Another parameter included is the learning rate that defines how far the line is shifted during each step, based on the information from the previous training step. These values are significant in the accuracy of the training model, and how long the training will take. For complicated models, initial conditions play a significant role in the determination of the outcome of training. Differences can be seen depending on whether a model starts off training with values initialized to zeroes versus some distribution of values. These parameters are commonly referred to as hyperparameters. The tuning of these parameters depends on the dataset, model, and the training process.

*7) Prediction*

ML is fundamentally using data to answer questions. Prediction is the final step where you get to answer few questions. This is the point where the value of ML is realized. The model gains independence from human interference and thus, draws its own conclusion, based on its data sets and training process. Here, eventually, the trained model can be used to predict the outcome for any desired inputs.

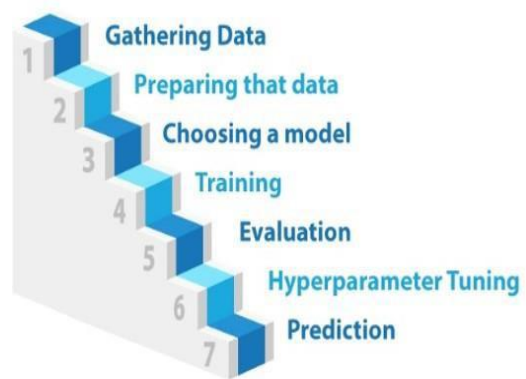

Figure 3. Seven steps of ML

## IV. SVM: BACKGROUND AND OVERVIEW

A SVM is a supervised learning algorithm that can use given data to solve certain problems by attempting to convert them into linearly separable problems [41]. SVM can be used for both classification and regression problems. It was first introduced by Vapnik [41-42] and was elaborated by Schölkopf et al. [43]. Although ANN is the most commonly used ML method for transient stability classification, it generally involves a broad training process and an intricate design procedure. Moreover, ANN usually performs well for interpolation but not so well for extrapolation, which reduces its generalization ability. They are more susceptible to becoming trapped in a local minimum. Although, majority of ML algorithms can overfit, if there is a dearth of training samples, but ANNs can also overfit if training goes on for a very long duration [42]. On the other hand, in the recent years, SVM classifiers have received a huge attention from power systems researchers because of producing single, optimum and automatic sparse solution by simultaneously minimizing both generalization and training error and unscrambling data by the large margin at high dimensional space [44-45]. Due to some of these downsides of ANN, it becomes essential to develop a more efficient classifier for transient stability status prediction. SVM does not suffer from these drawbacks and has the following advantages over ANN [11]: (1) less number of tuning parameters, (2) less susceptibility to overfitting, and (3) the complexity is dependent on number of support vectors (SVs) rather than dimensionality of transformed input space.

SVM classifiers depend on training points, which lie on the boundary of separation between different classes, where the evaluation of transient stability is important. A decent theoretical progress of the SVM, due to its basics built on the Statistical Learning Theory (SLT) [41], made it possible to develop fast training methods, even with large training sets and high input dimensions [46-48]. This useful characteristic can be applied to tackle the issue of high input dimension and large training datasets in the TSA problem. The basic implementation of an SVM, commonly known as a hard margin SVM, requires the binary classification problem to be linearly separable. This is frequently not the case in practical problems, and therefore, SVM provides a kernel trick to resolve this issue. The strength of the SVM algorithm is based on the use of this kernel trick to transform the input space into a higher dimensional feature space. This allows for defining a decision boundary that linearly separates the classes. The SVM algorithm attempts to find that decision boundary or hyperplane with the highest distance from each class [11].

## V. TYPES OF SVM

There are two main types of SVM [49-50]. They are described below.

*1)* Linear SVM

Linear SVM is used for data that are linearly separable i.e., for a dataset that can be categorized into two categories by utilizing a single straight line, as shown in Fig. 4. Such data points are termed as linearly separable data, and the classifier used is described as a linear SVM classifier.

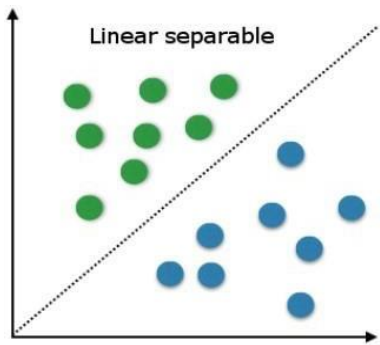

Figure 4. Linear SVM

*2)* Nonlinear SVM

Nonlinear SVM is used for data that are non-linearly separable data i.e., a straight line cannot be used to classify the dataset, as illustrated by Fig. 5. For this, something known as a kernel trick is used that sets data points in a higher dimension, where they can be separated using planes or other mathematical functions. Such data points are called non-linear data, and the classifier used is known as a nonlinear SVM classifier.

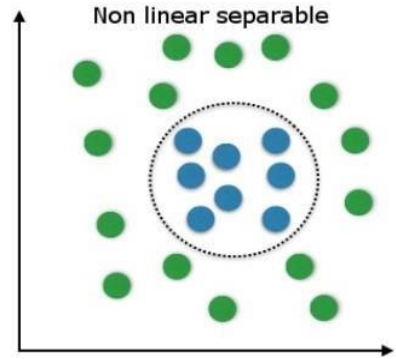

Figure 5. Nonlinear SVM

## VI. KERNEL FUNCTIONS IN SVM

In certain applications, the data set classes can be deeply overlapping, which makes it impossible to perform a linear classification in the feature space, even by introducing slack variables. The solution for these applications can be obtained by applying Cover's theorem [51]. It stipulates that it is highly probable to solve a nonlinear classification problem, using linear classifiers, by projecting the input set into a higher dimensional space using a nonlinear transformation function [51]. It is evident that the equation of the optimal hyperplane and the decision rule are function of the inner product of the SVs and the new input vector. By mapping the input set into a higher dimensional space, it is required to compute the high dimensional inner product of their transformation, which requires a good knowledge of the mapping function. According to the Hilbert-Schmidt theory for inner products in high dimensional spaces, computing $\langle \phi(x_i), \phi(x_j) \rangle$ is equivalent to computing a symmetric function $K(x_i, x_j)$ satisfying Mercer's theorem [52]. Here, $K$ is called the kernel function. Its main advantage is that it does not require any knowledge on the mapping function. Therefore, the use of the function $K$ is usually described as the kernel trick.

The kernel function is what is applied on each data instance to map the original non-linear observations into a higher-dimensional space in which they become separable. The kernel functions return the inner product between two points in a suitable feature space. The function of kernel is to take data as input and transform it into the required form using the transformation $\phi$ (as illustrated by Fig. 6). Commonly used kernel functions include the linear, polynomial, sigmoid, Gaussian Radial Basis Function (RBF), and Laplace RBF, as shown in Table II. The choice of the kernel function depends essentially on the data set and in certain cases several trials must be performed before choosing the appropriate one. The Gaussian kernel generally is preferred over others because it has the ability of mapping samples nonlinearly into a higher dimensional space, and therefore, unlike linear kernel, it can tackle the scenario when the relationship between class labels and attributes is nonlinear.

Although, sigmoid kernel performs like a Gaussian kernel for certain parameters, but there are some parameters for which the sigmoid kernel is not the dot product of two vectors, thus, it is invalid. Moreover, as compared to polynomial kernel, it has few hyperparameters (parameters whose values are used to control the learning process) [53].

To the best of author's knowledge, there exists no work which comprehensively reviews the research work related to SVM for TSA of power systems. Thus, the main objective and contribution of the current paper is to review major works related to ANN for TSA of power systems and provide research gaps and recommendations for future work.

TABLE II. COMMON SVM KERNELS

| Kernel | Equation |
|---|---|
| Polynomial | $K(x_i, x_j) = (x_i \cdot x_j)^{\text{degree}}$ |
| Gaussian RBF | $K(x_i, x_j) = e^{-\gamma \|x_i - x_j\|^2}$ |
| Linear | $K(x_i, x_j) = x_i \cdot x_j + \text{constant}$ |
| Laplace RBF | $K(x_i, x_j) = e^{-\frac{\|x_i - x_j\|}{\sigma}}$ |

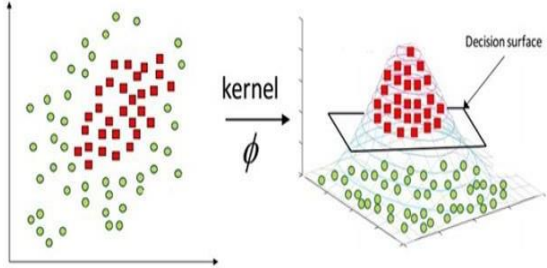

Figure 6. Illustration of kernel function

## VII. LITERATURE REVIEW: APPLICATION OF SVM FOR TSA

This section will review the application of SVM for problems involving TSA. Recently, SVM has been applied to power system transient stability classification problem. An SVM-based transient stability classifier was trained in [5] and its performance was compared with a Multi-layer Perceptron (MLP) classifier. Reference [54] devised a multiclass SVM classifier for TSA classification. Reference [55] recommended a SVM classifier to predict the transient stability status, using voltage variation trajectory templates. Reference [11] trained a binary SVM classifier, with combinatorial trajectories inputs, to assess the transient stability status. Reference [56] employed the SVM to rank the synchronous generators, based on transient stability severity, and consequently, classified them into vulnerable and nonvulnerable machines. Reference [57] proposed two SVMs, using Gaussian kernels, for classifying the post-fault transient stability status of the system. Reference [58] presented an SVM-based approach, for transient stability detection, using post-disturbance signals, from the optimally located distributed generations. Reference [59] proposed a multi-SVM power system TSA method, based on relief algorithm. Firstly, the suggested approach selected numerous feature subsets, with various size based on relief algorithm; then, used these chosen feature subsets for SVM training, and eventually, these trained SVMs were integrated to evaluate the transient stability of power system.

Reference [60] focused on the assessment of the transient stability of power systems, using pre-fault and fault duration data, measured by Wide Area Measurement System (WAMS). In the suggested approach, the time-synchronized values of voltage and current, created by synchronous generators, were measured using Phasor Measurement Units (PMUs), installed at generator buses, and given as input to the suggested algorithm, to obtain a proper feature set. Then, the devised feature set was applied to (SVM) classifier, to envisage the transient stability status. In [61], a different time series forecasting algorithm, using SVM, was proposed, which utilized synchronized phasor data, to provide fast transient stability swings prediction, for the use of emergency control. In [62], a conservative prediction model, for power system transient stability, was suggested, targeting at enhancing accuracy, for predicting the unstable cases. The model was recognized as an ensemble learning model, using multiple SVMs as sub-learning machines.

In [63], a twin convolutional SVM as supervised trajectory-based deep neural classifier was presented, which can remove the computational intricacy of kernel trick. The results demonstrated that the classification accuracy of the presented approach with a larger size window for each test systems exceeded 87% and it outperformed kernel-based approaches on test cases. In [64], an online power system TSA problem was mapped as a two-class classification problem and a novel ML algorithm known as the Core Vector Machine (CVM) was suggested to solve the problem based on PMUs big data. Compared with other SVMs, the devised CVM based assessment technique had the higher precision. Also, it had the least time consumption and space complexity. Based on the data collected from the PMU, a TSA method merging Stacked Automatic Encoder (SAE) and SVM was presented in [65]. Multi-layer abstract learning was performed on the original features by the SAE, and the extracted feature was used to train and test the SVM model.

Reference [66] proposed TSA of a large practical power system using two ANN approaches: Probabilistic Neural Network (PNN) and Least Squares SVM (LS-SVM). Transient stability of the power system was first evaluated based on the generator relative rotor angles (obtained using time domain simulations). Classification results demonstrated that the PNN gives faster, and more accurate results for TSA when compared to the LS-SVM. Considering the fact that the traditional SVM method cannot avoid false classification, [67] suggested a novel approach to solve the weaknesses of traditional SVM, which can enhance the interpretability of results, and avoid the problem of false alarms and missed alarms. In this approach, two enhanced SVMs, known as the Aggressive SVM (ASVM) and Conservative SVM (CSVM), were presented to increase the accuracy of the classification. Cases studies on IEEE 39-bus system and a real provincial power network illustrated the efficacy and viability of the suggested technique.

In [68], a unique TSA algorithm was presented, where SVMs were employed as pattern classifiers. SVMs with different kernel functions and kernel parameters were constructed and trained to compute hyperplanes that split the stable and unstable states of power system for $(n-1)$ faults. The simulation results obtained using three benchmark systems demonstrated the good capacity for fuzzy combined SVM classifiers in TSA. Compared with traditional SVM, [69] devised an advanced TSA system using Multi-layer SVM (ML-SVM) approach. In the proposed method, a Genetic Algorithm (GA) was used in ML-SVM to identify the valued feature subsets with differing numbers of feature. Transient stability of the system was determined based on the generator relative rotor angles. The simulation results demonstrated that the presented approach could lessen the likelihood of misclassification. Reference [70] proposed a comparative analysis of two different machine learning (ML) algorithms, i.e., ANN and SVM, for online transient stability prediction, considering various uncertainties, such as load, network topology, fault type, fault location, and Fault Clearing Time (FCT). The results for the IEEE 14-bus system demonstrated that both ANN and SVM can rapidly estimate the transient stability; however, ANN outclassed SVM as its classification performance and computational performance were established to be greater.

Reference [71] studied the application of a binary SVM-based supervised ML, for forecasting the transient stability status of a power system, considering uncertainties of various factors, such as load, faulted line, fault type, fault location and FCT. The SVM was trained using a Gaussian RBF kernel and its hyperparameters were optimized using Bayesian optimization. Results obtained for the IEEE 14-bus test system demonstrated that the suggested procedure offers a rapid approach for Probabilistic Transient Stability (PTS) status prediction, with an excellent accuracy. Reference [72] presented an online assessment approach that may be applied in TSA forecasting which can predict the future behavior by using Support Vector Regression (SVR) and the incoming data from PMUs. The presented strategy was based on the Self-Organizing Feature Map (SOFM) that can locate the similar training input data and cluster them into several classes. Forecasting results of simulation on IEEE 39-bus system demonstrated the viability of this model. In [73], SVMs were solved by the 2nd order convex programming and the ultimate solution of SVMs obtained was unique and optimal. Experiments substantiate the dominance of presented strategy applied for TSA in power systems by comparing with Back Propagation (BP) approach and RBF. Reference [74] proposed TSA using LS-SVM and principle component analysis. The data collected from the time domain simulations was used as inputs to the LS-SVM in which LS-SVM was used as a classifier to establish the stability state of a power system. To validate the efficacy of the suggested LS-SVM method, its performance was compared with the MLP neural network. Results demonstrated that the LS-SVM gives faster, and more accurate TSA compared to the MLP neural network in terms of classification results.

Reference [75] presented an enhanced SVM method for TSA of power system. Firstly, the initial feature set was computed by a simple calculation of the initial operating parameters of the system, such as projection energy function feature. Consequently, the feature sets were used for the TSA problem of SVM with pinball loss. The viability and rationality of the presented strategy was shown using the IEEE 145-bus system and a real power system. Reference [76] presented a parameter searching method based on Bayesian optimization that incorporates verification curves and Recursive Feature Elimination (RFE) to optimize the SVM for TSA. The prior function in Bayesian optimization was used to simulate the distribution of hyperparameters. Also, the acquisition function was used to determine the best search point, thus significantly decreasing the computational burden of offline training. The simulation results on IEEE 39-bus system showed that the offline training time of the SVM based on the enhanced Bayesian optimization was considerably curtailed. Some other appropriate work dealing with SVM-based TSA can be found in [77, 78-84].

## VIII. RESEARCH GAPS AND FUTURE RECOMMENDATIONS

Based on the detailed literature review and to the best of author's knowledge, there exists no research work on PTS which uses SVM-based ML approach, considering the uncertainties of load, faulted line, fault type, fault location (on the line), and FCT. Moreover, [85] specifically mentions the potential of SVM for online TSA. In addition, [86-90] strongly indicate that ML is a promising and upcoming approach for online Dynamic Security Assessment (DSA). Thus, one of the main research gap is to predict PTS status using an SVM-based ML approach.

Also, there is a dearth of work which incorporates renewable energy in TSA prediction using SVM. As the amount of renewable energy is increasing continually in the power system, the dynamics of power system are becoming more intricate. More comprehensive studies and simulations are required to understand the behavior of the system under renewable energy integration. Network topology changes are often ignored in the existing research work. It is significant to train the SVM model for changing topologies. In this regard, the network base topology can be investigated towards accomplishing an optimized network topology from the standpoint of transient stability of the whole system. Novel techniques must be researched and applied such that the SVM model can adapt to any topology of the power system.

Currently, SVM-based TSA methods face some challenges. First, it is very difficult to obtain large-scale, balanced data of random input variables with accurate labels in real-world scenarios. On top of that, existing SVM-based TSA methods act as a black box which have poor interpretability, which also limits their application in actual power systems. Approaches must be formulated to integrate various kinds of stability into a single index and consequently, utilize SVM to predict its value. It is highly recommended that future research work on SVM based approach to power system transient stability should focus on comprehensive authentication of the approach using large scale test system which have similar attributes of uncertainty and randomness as that of a modern power system. The simulations results obtained for SVM accuracy and performance must be tested using a Real Time Digital Simulator (RTDS). To accomplish this, the power system with wide area measurement can be simulated on the RTDS and the signals can be sampled from the input-output port of RTDS and fed to the predictor to evaluate its performance [55].

In future studies, it is also recommended to incorporate unobserved real-time operating conditions of power networks such as information lost due to the communication failure (unavailability) and absence of quality of power system dynamic responses (noisy data). As different feature subsets have different useful information of a power system; therefore, comprehensive use of this information for TSA must be made. Misclassification and missed classification have entirely dissimilar impact for the stability of the system. Therefore, using different feature subsets to train SVMs and consequently, integrating the result can considerably reduce the misclassified samples, which is of great significance for TSA in practical power systems. Assessment of various procedures to mitigate effects of imbalance datasets is also significant in this regard. Existing research works propose training the SVM models offline based on the historical data which could be applicable in real-time. However, the next imperative step in this direction could be to use reinforcement learning for power system transient stability surveillance which can train the SVM models online simultaneously with the existing data while also examining the network at the same time.

The performance of the SVM-based ML approach depends on the quality and quantity of the data. In power systems, the required data are either unavailable, unlabeled or have low quality. Hence, the necessary information is collected using equivalent system models. However, the design of the database generation process can produce biased models, which can cause an overestimation of its performance during assessment. Consequently, it is essential to research further into their robustness and reliability [91].

In addition to the above-mentioned recommendations and research gaps, and intensifying the adoption of ML in the power system, it is imperative that the whole power community collaborate on laying a strong foundation for the continuous growth of ML. The major components to form this foundation involve but not restricted to public accessible benchmark datasets, open source developing environment, standardized testing configurations, etc. An open and standardized research environment can hasten the conversion of technical accomplishments to the practical applications [92].

The present study provided a review of some major research works and potential future research avenues associated with SVM application to TSA. This can be a remarkable offset for researchers in the domain of ML, power system stability and operation, particularly in the presence of uncertainty. Recent research [93-103] reveals that there is a lot of scope in this domain, and its potential must be fully investigated.

## IX. CONCLUSION AND FUTURE WORK

TSA of the power system is a critical issue with escalating demands and numerous operating restrictions. With the increasing uncertainty, renewable energy generation, and electricity market deregulation, its accurate evaluation cannot be overestimated. The constraints of online TSA for modern power systems have become quite strict. Moreover, inability to fulfil these requirements can cause instability which can result in cascading outages and blackouts, hence, causing economic, social, and technical losses. Novel soft computing approaches based on ML, such as SVM, can play a valuable part in ensuring that these requirements are met. Therefore, this paper provided a review of works related to application of SVM to TSA. It is believed that this review will provide a good basis for researchers in the field of SVM and power system transient stability, and consequently, help them understand the existing research status and questions. As a future work, numerous reviews can be conducted using other ML approaches and a comparative analysis can be drawn. Moreover, ensemble ML approaches can be explored for TSA which combine two or more ML approaches to achieve better performance. Merging quantum computing with ML for TSA is another open area of research.